\documentclass[twocolumn]{aastex63}

\newcommand{\mearth}{$M_{\oplus}$}

\newcommand{\hei}{He\,I}

\received{}
\revised{}
\accepted{\today}
\submitjournal{ApJL}

\shorttitle{Transit Spectroscopy for AU Mic b}
\shortauthors{Hirano et al.}
\graphicspath{{./}{figures/}}
\begin{document}

\title{Limits on the Spin-Orbit Angle and Atmospheric Escape for the 22 Myr-old Planet AU Mic b\footnote{Based on data collected at Subaru Telescope, which is operated by the National Astronomical Observatory of Japan, and the W. M. Keck Observatory, which is supported by the W. M. Keck Foundation and operated by the California Institute of Technology, the University of California and the National Aeronautics and Space Administration.}}

\correspondingauthor{Teruyuki Hirano}
\email{hirano@geo.titech.ac.jp}

%\author[0000-0002-0786-7307]{Greg J. Schwarz}
%\affiliation{American Astronomical Society \\
%1667 K Street NW, Suite 800 \\
%Washington, DC 20006, USA}

\author[0000-0003-3618-7535]{Teruyuki Hirano}
\affiliation{Department of Earth and Planetary Sciences, Tokyo Institute of Technology, 2-12-1 Ookayama, Meguro-ku, Tokyo 152-8551, Japan}

\author[0000-0003-2310-9415]{Vigneshwaran Krishnamurthy}
\affiliation{Department of Earth and Planetary Sciences, Tokyo Institute of Technology, 2-12-1 Ookayama, Meguro-ku, Tokyo 152-8551, Japan}

\author[0000-0002-5258-6846]{Eric Gaidos}
\affiliation{Department of Earth Sciences, University of Hawai'i at M\={a}noa, Honolulu, HI 96822, USA}

\author[0000-0002-1050-4056]{Heather Flewelling}
\affiliation{Institute for Astronomy, University of Hawaii at M\={a}noa, Honolulu, HI 96822, USA}

\author[0000-0003-3654-1602]{Andrew W. Mann}
\affiliation{Department of Physics \& Astronomy, University of North Carolina at Chapel Hill, Chapel Hill, NC 27559, USA}

\author[0000-0001-8511-2981]{Norio Narita}
\affiliation{Komaba Institute for Science, The University of Tokyo, 3-8-1 Komaba, Meguro, Tokyo 153-8902, Japan}
\affiliation{JST, PRESTO, 3-8-1 Komaba, Meguro, Tokyo 153-8902, Japan}
\affiliation{Astrobiology Center, NINS, 2-21-1 Osawa, Mitaka, Tokyo 181-8588, Japan}
\affiliation{National Astronomical Observatory of Japan, NINS, 2-21-1 Osawa, Mitaka, Tokyo 181-8588, Japan}
\affiliation{Instituto de Astrof\'{i}sica de Canarias (IAC), 38205 La Laguna, Tenerife, Spain}

%\author{TBD}

\author[0000-0002-8864-1667]{Peter Plavchan}
\affiliation{Department of Physics and Astronomy, George Mason University, 4400 University Drive, MSN 3F3, Fairfax, VA 22030, USA}

\author{Takayuki Kotani}
\affiliation{Astrobiology Center, NINS, 2-21-1 Osawa, Mitaka, Tokyo 181-8588, Japan}
\affiliation{National Astronomical Observatory of Japan, NINS, 2-21-1 Osawa, Mitaka, Tokyo 181-8588, Japan}
\affiliation{Department of Astronomy, School of Science, The Graduate University for Advanced Studies (SOKENDAI), 2-21-1 Osawa, Mitaka, Tokyo, Japan}

\author[0000-0002-6510-0681]{Motohide Tamura}
\affiliation{Astrobiology Center, NINS, 2-21-1 Osawa, Mitaka, Tokyo 181-8588, Japan}
\affiliation{National Astronomical Observatory of Japan, NINS, 2-21-1 Osawa, Mitaka, Tokyo 181-8588, Japan}
\affiliation{Department of Astronomy, Graduate School of Science, The University of Tokyo, 7-3-1 Hongo, Bunkyo-ku, Tokyo 113-0033, Japan}

\author[0000-0002-6197-5544]{Hiroki Harakawa}
\affiliation{Subaru Telescope, 650 N. Aohoku Place, Hilo, HI 96720, USA}

\author[0000-0003-0786-2140]{Klaus Hodapp}
\affiliation{University of Hawaii, Institute for Astronomy, 640 N. Aohoku Place, Hilo, HI 96720, USA}

\author{Masato Ishizuka}
\affiliation{Department of Astronomy, Graduate School of Science, The University of Tokyo, 7-3-1 Hongo, Bunkyo-ku, Tokyo 113-0033, Japan}

\author{Shane Jacobson}
\affiliation{University of Hawaii, Institute for Astronomy, 640 N. Aohoku Place, Hilo, HI 96720, USA}

\author[0000-0003-0114-0542]{Mihoko Konishi}
\affiliation{Faculty of Science and Technology, Oita University, 700 Dannoharu, Oita 870-1192, Japan}

\author[0000-0002-9294-1793]{Tomoyuki Kudo}
\affiliation{Subaru Telescope, 650 N. Aohoku Place, Hilo, HI 96720, USA}

\author{Takashi Kurokawa}
\affiliation{Astrobiology Center, NINS, 2-21-1 Osawa, Mitaka, Tokyo 181-8588, Japan}
\affiliation{Institute of Engineering, Tokyo University of Agriculture and Technology, 2-24-16, Nakacho, Koganei, Tokyo, 184-8588, Japan}
%\affiliation{National Astronomical Observatory of Japan, NINS, 2-21-1 Osawa, Mitaka, Tokyo 181-8588, Japan}

\author[0000-0002-4677-9182]{Masayuki Kuzuhara}
\affiliation{Astrobiology Center, NINS, 2-21-1 Osawa, Mitaka, Tokyo 181-8588, Japan}
\affiliation{National Astronomical Observatory of Japan, NINS, 2-21-1 Osawa, Mitaka, Tokyo 181-8588, Japan}

\author{Jun Nishikawa}
\affiliation{National Astronomical Observatory of Japan, NINS, 2-21-1 Osawa, Mitaka, Tokyo 181-8588, Japan}
\affiliation{Department of Astronomy, School of Science, The Graduate University for Advanced Studies (SOKENDAI), 2-21-1 Osawa, Mitaka, Tokyo, Japan}
\affiliation{Astrobiology Center, NINS, 2-21-1 Osawa, Mitaka, Tokyo 181-8588, Japan}

\author{Masashi Omiya}
\affiliation{Astrobiology Center, NINS, 2-21-1 Osawa, Mitaka, Tokyo 181-8588, Japan}
\affiliation{National Astronomical Observatory of Japan, NINS, 2-21-1 Osawa, Mitaka, Tokyo 181-8588, Japan}

\author{Takuma Serizawa}
\affiliation{Institute of Engineering, Tokyo University of Agriculture and Technology, 2-24-16, Nakacho, Koganei, Tokyo, 184-8588, Japan}

\author{Akitoshi Ueda}
\affiliation{National Astronomical Observatory of Japan, NINS, 2-21-1 Osawa, Mitaka, Tokyo 181-8588, Japan}

\author[0000-0003-4018-2569]{S\'ebastien Vievard}
\affiliation{Astrobiology Center, NINS, 2-21-1 Osawa, Mitaka, Tokyo 181-8588, Japan}
\affiliation{Subaru Telescope, 650 N. Aohoku Place, Hilo, HI 96720, USA}

%\nocollaboration{2}

%% Note that the \and command from previous versions of AASTeX is now
%% depreciated in this version as it is no longer necessary. AASTeX 
%% automatically takes care of all commas and "and"s between authors names.

%% AASTeX 6.3 has the new \collaboration and \nocollaboration commands to
%% provide the collaboration status of a group of authors. These commands 
%% can be used either before or after the list of corresponding authors. The
%% argument for \collaboration is the collaboration identifier. Authors are
%% encouraged to surround collaboration identifiers with ()s. The 
%% \nocollaboration command takes no argument and exists to indicate that
%% the nearby authors are not part of surrounding collaborations.

%% Mark off the abstract in the ``abstract'' environment. 
\begin{abstract}
We obtained spectra of the pre-main sequence star AU Microscopii during a transit of its Neptune-sized planet to investigate its orbit and atmosphere.  We used the high-dispersion near-infrared spectrograph IRD on the Subaru telescope to detect the Doppler ``shadow" from the planet and constrain the projected stellar obliquity.
%we find that the stellar spin and orbit are aligned ($\lambda=-4.7_{-6.4}^{+6.8}$ degrees), suggesting that the planet formed and possibly migrated within the protoplanetary disk. 
Modeling of the observed planetary Doppler shadow suggests a spin-orbit alignment of the system ($\lambda=-4.7_{-6.4}^{+6.8}$ degrees), but additional observations are needed to confirm this finding.
We use both the IRD data and spectra obtained with NIRSPEC on Keck-II to search for absorption in the 1083\,nm line of metastable triplet He\,I by the planet's atmosphere and place an upper limit for the equivalent width of 3.7 m\AA{} at 99\,\% confidence. With this limit and a Parker wind model we constrain the escape rate from the atmosphere to $<0.15-0.45$ \mearth\,Gyr$^{-1}$, comparable to the rates predicted by an XUV energy-limited escape calculation and hydrodynamic models, but refinement of the planet mass is needed for rigorous tests. 
\end{abstract}

%% Keywords should appear after the \end{abstract} command. 
%% See the online documentation for the full list of available subject
%% keywords and the rules for their use.

%% From the front matter, we move on to the body of the paper.
%% Sections are demarcated by \section and \subsection, respectively.
%% Observe the use of the LaTeX \label
%% command after the \subsection to give a symbolic KEY to the
%% subsection for cross-referencing in a \ref command.
%% You can use LaTeX's \ref and \label commands to keep track of
%% cross-references to sections, equations, tables, and figures.
%% That way, if you change the order of any elements, LaTeX will
%% automatically renumber them.
%%
%% We recommend that authors also use the natbib \citep
%% and \citet commands to identify citations.  The citations are
%% tied to the reference list via symbolic KEYs. The KEY corresponds
%% to the KEY in the \bibitem in the reference list below. 

%%%%%%%%%%%%%%%%%%%%%%%%%%
\section{Introduction} \label{sec:intro}

Detection and characterization of exoplanets in stellar clusters, young moving groups, and even younger star-forming regions can be used to test models of planet formation and evolution, including cooling and contraction \citep[e.g.,][]{Vazan2018}, loss of light-element atmospheres \citep[e.g.,][]{Ginzburg2018,Owen2019}, and orbital evolution \citep[e.g,][]{Spalding2016}.   Such planets have been discovered using the radial velocity (RV) technique \citep[e.g.,][]{2007ApJ...661..527S, 2012ApJ...756L..33Q} and more recently, with the photometric (transit) method by the {\it K2} mission \citep[e.g.,][]{David2016,2016ApJ...818...46M,2016AJ....152...61M,Mann2018,David2019}. 
The {\it TESS} mission is now surveying the entire sky for transiting planets, finding systems with host stars that span a greater range of ages (including moving group members), and they are closer and brighter and hence more amenable to characterization \citep[e.g.,][]{2019ApJ...880L..17N,Mann2020}.

Using {\it TESS} photometry, \citet{2020Natur.582..497P} discovered a Neptune-sized planet ($R_p= 4.3R_{\oplus}$) on a 8.5-day transiting orbit around the M star AU Mic, a member of the 
$\approx20$\,Myr-old $\beta$ Pictoris moving group \citep{Mamajek2014} and the second nearest known pre-main sequence star.  The star possesses a well-studied debris disk \citep[e.g.,][]{Grady2020}.   
%The planet has a radius $R_p= 4.3R_{\oplus}$, and its mass is constrained by RVs to $<58 M_{\oplus}$ at 99\% confidence (P. Plavchan, pers. comm.).  
Given the proximity (9.7\,pc) and brightness ($V=8.6$) of its host star, AU Mic b is an unparalleled  opportunity to study the properties of an infant %adolescent 
planet, particularly by spectroscopic observations during transits. 

We observed AU Mic during a single transit of ``b" to constrain the planet's orbit and detect any (escaping) atmosphere.  The stellar obliquity with respect to the planetary orbit can be measured using the Rossiter-McLaughlin (RM) effect \citep[e.g.,][]{2005ApJ...631.1215W}.  This obliquity reflects the dynamical history of a planetary system; a measurement for a very young system offers more leverage to test different scenarios, including planet migration and planet-planet scattering \citep[e.g.,][]{2012ARA&A..50..211K,  2007ApJ...669.1298F, 2011ApJ...742...72N}.  A planet's extended or escaping atmosphere can be detected via transit-associated absorption.  The infrared (1083\,nm) lines of metastable ``triplet" \hei\ are accessible from the ground.   The escape of H/He-rich envelopes could be responsible for a``Neptune desert" \citep{2016NatCo...711201L} and a ``radius valley" \citep[e.g.,][]{2017AJ....154..109F} found in the exoplanet radius-period distribution. The proliferation of RV-capable echelle spectrographs operating in the near-infrared has made simultaneous observations of RM and the He I line possible, as we demonstrate here. 

%%%%%%%%%%%%%%%%%%%%%%%%%%
\section{Observations and Data Reduction} \label{sec:obs}
\subsection{Subaru/IRD}
On UT 2019 June 17, a transit of AU Mic b was observed with the InfraRed Doppler (IRD) spectrograph \citep[$\lambda/\Delta\lambda\approx 70,000$, $\lambda=930-1740$ nm:][]{2012SPIE.8446E..1TT, 2018SPIE10702E..11K} 
on the Subaru telescope. %on Maunakea.  
%Using the AO188 adaptive-optics system \citep{2008SPIE.7015E..10H}, 
%We injected the stellar light into the spectrograph, as well as the simultaneous reference light from the laser-frequency comb \citep[LFC:][]{2016SPIE.9912E..1RK}. 
Since AU Mic was too low in elevation when the transit ingress started, we missed the first 15 min of the transit. We continuously observed the target with individual integrations of 60 sec for about 3.8 hours covering the rest of the transit. 
%as well as some time after transit egress.  
We suspended the observation of AU Mic about 40 min after the egress, and obtained  spectra of the A0 star HIP 98926 at a similar airmass for telluric correction.  After taking several spectra of HIP 98926, we resumed the observation of AU Mic until the end of the night.  
%Since IRD's RV analysis pipeline requires the target's spectra taken on different epochs to disentangle stellar lines from the telluric ones, we observed AU Mic with IRD in 2019 June and October. Those spectra are used to extract a telluric-free stellar template spectrum for RV measurements. 
%Raw IRD data were reduced using {\tt IRAF} software as well as our custom codes
%\citep{2018SPIE10702E..60K}. 
As explained in \citet{2018SPIE10702E..60K}, raw IRD data were reduced using {\tt IRAF} software as well as our custom codes. 
%to analyze the bias and readout noise of the detectors \citep{2018SPIE10702E..60K}. 
%After extracting the one-dimensional (1D) spectra, wavelengths were calibrated with frames in which LFC light was injected into both stellar and reference fibers.  
The extracted 1-D spectra have a per-pixel signal-to-noise ratio (S/N) of 100-120 at 1000 nm. 

\subsection{Keck-2/NIRSPEC}

$Y$-band (947-1121\,nm) spectra of AU Mic were obtained with NIRSPEC \citep{McLean1998} on the Keck-II telescope on Maunakea.  The high-resolution echelle mode with a $0\farcs288$ (3-pixel) $\times$ $12^{\prime\prime}$ slit delivers $\lambda/\Delta \lambda  \approx 37,500$.  The star was observed from UT 2019 June 17 %10:04:45
$\mathrm{MJD}=58651.419967$, after ingress, at an airmass of 2.95, %until UT 14:27:58
until $\mathrm{MJD}=58651.602753$, well after egress and when the star was at an airmass of 1.65.  AB nodding was performed for sky subtraction, with the A and B beams separated along the slit by $6\farcs3$ or 44 pixels.  The  integration time per beam was 59 sec and readout used MCDS (four-read double-sampling).  194 integrations were obtained but only 94 A-B beam pairs were usable.   The A0 star HD 152849 was observed for telluric correction immediately before the start of AU Mic observations.  Flat fields and darks were obtained with the same integration time (1.47 sec) and a master flat was produced by median combination.  Per pixel S/N ratio in the vicinity of the \hei\ line is 450-2200 per integration, with a median of 1700.  A-B beam image pairs were differenced, flattened, and the orders extracted using custom {\tt Python} scripts.  The \hei\ line appears in two adjacent orders (70 and 71) but 71 suffers from lower SNR and artifacts near the array edge and only 70 was used.         

\subsection{UH-2.2m/SNIFS}

AU Mic was simultaneously observed with the Super-Nova Integral Field Spectrograph (SNIFS) on the UH 2.2-m telescope on Maunakea.  SNIFS uses a lenslet array to re-image a target onto gratings in separate blue (320-520\,nm) and red (510-870\,nm) channels with $\lambda/\Delta\lambda \approx 900$ \citep{Lantz2004}.   Per pixel, per integration S/N ratio was about 650, 210 and 145 in the H I Balmer $\alpha$, $\beta$, and $\gamma$ lines, respectively. Spectra were extracted and wavelength calibrated using flat-fields and arcs taken at the same pointing.  71 spectra of AU Mic were obtained over 4 hours %5 min 
between $\mathrm{MJD}=58651.4367$ and 58651.6064.  Spectra of telluric calibration stars (HR 4468, GD 153, BD+332642, HR 7596, HR 7950) were obtained before and after AU Mic.  Spectra were extracted by the automatic SNIFS pipeline and equivalent widths of H$\alpha$, $\beta$, and $\gamma$ lines were measured in the intervals 655.4-657.5, 485.4-486.9, and 433.2-434.8 nm, respectively, with the continuum obtained from adjacent $\sim$1\,nm-wide intervals.  

%%%%%%%%%%%%%%%%%%%%%%%%%%
\section{Analyses and Results} \label{sec:ana}

\subsection{RV Analysis and Modeling of the RM Effect} \label{sec:RV}

To estimate the stellar obliquity, we first analyzed the effective RVs during the transit. Putting the 1D IRD spectra into the RV pipeline, we measured relative RVs for AU Mic (Appendix \ref{sec:app1}). 
%Our RV-analysis pipeline is described in Hirano et al. (2020). 
%In brief, extracting the instantaneous instrumental profile (IP) of the spectrograph for each frame from the LFC spectrum, the pipeline models and fits a number of small spectral segments by the forward modeling technique. 
%The resulting RVs during the transit night are presented in Appendix \ref{sec:app1}. %Figure \ref{fig:RV}.  
%The typical internal RV error during the night was $8-10$ m s$^{-1}$. 
The RV data exhibit a bump around the expected egress, but the lack of ingress and baseline RVs prevented us from evaluating the obliquity. To constrain the projected stellar obliquity $\lambda$, we attempted to fit the observed RVs. 
%based on the transit ephemeris and system parameters reported in Plavchan et al. (in press). 
After some attempts to fit the observed RVs with different subsets of data and different assumptions on the baseline and priors, however, 
we found that we are unable to obtain a reliable estimate for $\lambda$ from our RV data set, mostly due to the lack of
out-of-transit RV data and unknown behavior of activity-induced RV variations (Appendix \ref{sec:app1}). 
Hence, we decided to determine $\lambda$ using an alternative technique, Doppler-shadow measurements.

\subsection{Doppler-shadow Analysis} \label{sec:DT}

Partial occultation of the rotating stellar disk by a transit is manifested as
a distortion in the stellar line \citep[e.g.,][]{2010ApJ...709..458H}.  Since the cross-correlation function (CCF) of an observed spectra against a template reflects the instantaneous line profile, the time sequence of the CCF residual from the mean out-of-transit CCF exhibits a time-varying anomaly representing a planet's Doppler ``shadow"  and this technique has been used to measure the obliquity of host stars
\citep[e.g.,][]{2010MNRAS.407..507C, 2017AJ....154..137J}.

%%%%%%%%%%%%%%%%%%%
\begin{figure*}
\centering
\includegraphics[width=18cm]{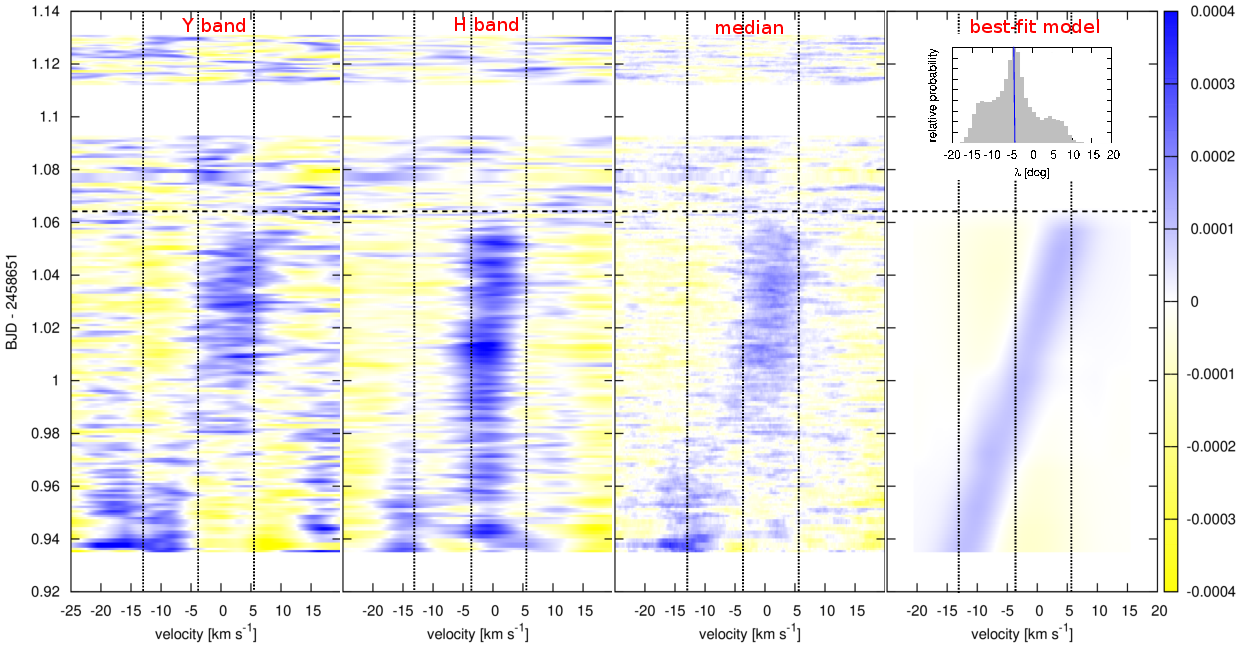}
\caption{Residual CCF maps for IRD spectra. 
The left two panels illustrate the combined CCFs for the $Y-$band 
orders (leftmost) and $H-$band orders (second left), respectively. 
The $H-$band spectra are more vulnerable to telluric lines, and affected
by the detector's persistence. For reference, we obtained 
$\lambda=7.7_{-4.6}^{+4.7}$ degrees and $\lambda=85.1_{-1.1}^{+1.4}$ degrees
for the $Y-$band and $H-$band maps, respectively, but those estimates are
severely affected by the systematic CCF noise. 
%Those CCF maps are contaminated by order-dependent, low-frequency CCF modulations 
%likely due to variations in the blaze shape, telluric lines, and/or the detector's
%persistence. 
The third panel from the left displays the median-combined residual CCF map 
using all available orders. 
%, in which the CCF modulations are suppressed and the planet shadow is seen more clearly. 
The rightmost panel is the best-fit model for the median-combined CCF map, 
and its inset displays the posterior distribution of $\lambda$ for the MCMC fit. 
For each panel, the three vertical dotted lines indicate the CCF line center (middle),
representing the barycentic RV of AU Mic, and $\pm v\sin i$ from the center. 
The horizontal dashed line represents the transit end time. 
}
\label{fig:DT}
\end{figure*}
%%%%%%%%%%%%%%%%%%%
We performed the Doppler-shadow analysis for the IRD spectra, 
following \citet{2020ApJ...890L..27H}. 
In computing the CCF, %for each IRD frame, 
we used a template spectrum representing
an early M star, generated from observed IRD spectra of GJ 436 (Hirano et al. submitted). 
For each frame, the spectrum for each echelle order was cross-correlated against the 
template, and we normalized the resulting CCF so that the pseudo-continuum of the
CCF wings becomes unity. 
In \citet{2020ApJ...890L..27H}, we summed the CCFs for individual orders before 
the normalization so that the combined CCF conserves the S/N information
for each order. However, we found that a similar analysis for AU Mic's spectra resulted in a low-frequency CCF modulation in time, which differs in each order (left two panels of Figure \ref{fig:DT}). This sort of 
modulation was previously seen in our analysis 
\citep{2020ApJ...890L..27H, Gaidos2020}, and
which mostly originates from a gradual variation of the blaze function, 
time-varying telluric lines, and/or the detector's persistence. 
These modulations could be removed by e.g., fitting
a low-order polynomial to the out-of-transit CCF data
and interpolating the variation during the transit. 
For AU Mic's IRD data, however, we lack the ingress and baseline
spectra before the transit, and we are unable to remove those 
modulations by such processing. 

To remove the order-dependent CCF modulation, we first 
subtracted the mean out-of-transit CCF from the individual CCF for 
each order. We then combined the residual CCFs for all orders by median
to obtain the final residual CCFs. This way of combining CCFs does
not properly reflect the S/N information for individual orders, but we found through
our experience that this processing can suppress the low-frequency modulation 
in the resulting residual CCF.

The third panel from the left in Figure \ref{fig:DT} depicts this extracted residual 
CCF as a function of time, after Doppler-shifting each frame by the 
barycentric motion of Earth. %and aligning all frames in the common reference frame. 
The planet shadow, manifested as a CCF bump,
appears to move from the blueshifted part of the profile to the redshifted side and 
disappears at the expected transit end time ($\mathrm{BJD}- 2458651\approx1.06$). 
In order to model this residual CCFs, we computed a number of theoretical CCFs
following \citet{2020ApJ...890L..27H}. %using GJ 436's spectrum broadened by the 
%rotation plus macroturbulence broadening kernel 
%($5\,\mathrm{km~s}^{-1}\leq v\sin i\leq 15$ km s$^{-1}$ for the rotation velocity), 
We created mock IRD spectra during the transit for different planet positions
on the stellar disk, and put those spectra into the CCF calculations. 
Based on these theoretical CCFs, the CCF residual map against time is 
generated by interpolations for any given set of $\lambda$, $v\sin i$, and the other 
parameters for AU Mic b.

%%%%%%%%%%%%%%%%%%%%%%%%%%%%%%%%%%%%%%%%%%%%%%%%%%%%%%%%%%%%%%%%%%%%%%
\begin{table}[t]
%\tabletypesize{\small}
\centering
\caption{Derived Parameters for the Doppler-Shadow Analysis. The symbols $\mathcal{U}$ 
and $\mathcal{N}$ represent the uniform and Gaussian priors, respectively.}\label{hyo1}
\begin{tabular}{lcc}
\hline\hline
Parameter & Value & Prior \\\hline
\multicolumn{2}{l}{\it (derived parameters)} & \\
$\lambda$ (degrees) & $-4.7_{-6.4}^{+6.8} $ & $\mathcal{U}\,[-180, +180]$ \\
$v\sin i$ (km s$^{-1}$) & $9.23_{-0.31}^{+0.79} $ & $\mathcal{U}\,[5, 15]$ \\
$b$ & $0.18_{-0.03}^{+0.07} $ & $\mathcal{U}\,[0, \infty]$ \\
$a/R_s$ & $19.34_{-0.59}^{+0.45}$ & $\mathcal{N}(19.1, 1.7)$\\
$T_c$ (BJD-2458651) & $0.99275 \pm 0.00061$ & $\mathcal{N}(0.99351, 0.00070)$\\\hline
\multicolumn{2}{l}{\it (fixed parameters)} &\\
$R_p/Rs$ & 0.0514 & \\
$e$ & 0 & \\
$P$ (days) & 8.46321 & \\
\hline
\end{tabular}
\end{table}
%%%%%%%%%%%%%%%%%%%%%%%%%%%%%%%%%%%%%%%%%%%%%%%%%%%%%%%%%%%%%%%%%%%%%%

We implemented a Markov Chain Monte Carlo (MCMC) analysis to estimate $\lambda$ using the CCF model above. We let $\lambda$, $v\sin i$, and the transit impact parameter $b$ float with uniform priors, and allowed the scaled semi-major axis $a/R_s$ and mid-transit time $T_c$ to vary with Gaussian priors (Table \ref{hyo1}). 
%($a/Rs=19.1\pm1.7$ and $T_c=2458651.99351\pm0.00070$).  %The other parameters (e.g., the orbital period $P$ and planet-to-star radius ratio $R_p/R_s$) were held fixed at the values in the discovery paper. 
The result of the analysis is summarized in Table \ref{hyo1}. 
The reduced $\chi^2$ for the best-fit model is 0.87. 
The best-fit obliquity ($\lambda=-4.7_{-6.4}^{+6.8}$ degrees) implies a good spin-orbit alignment for the AU Mic system. 
%The estimated $v\sin i$ of $9.23_{-0.31}^{+0.79} $ km s$^{-1}$ is compatible with the 
%result of the line analysis (Appendix \ref{sec:app1}). 

The estimated rotation velocity ($v\sin i=9.23_{-0.31}^{+0.79} $ km s$^{-1}$) is consistent within $\approx 1.5\,\sigma$ with the spectroscopic value ($8.7\pm0.2$ km s$^{-1}$) reported in \citet{2020Natur.582..497P}.  
%From the width of the CCF profile, we also obtained a similar estimate for $v\sin i$ ($\approx 8.4$ km s$^{-1}$; see Appendix \ref{sec:app1}).  
However, those estimates are slightly larger than the expected rotation velocity at the stellar equator ($2-3\,\sigma$ level), inferred from the stellar radius and rotation period based on the {\it TESS} photometry ($v_\mathrm{eq}=7.81\pm0.31$ km s$^{-1}$).  The reason for the disagreement is not known, but it suggests that (1) both spectroscopic measurements and Doppler-shadow analysis have relatively large systematic errors in $v\sin i$, (2) the stellar radius reported in \citet{2020Natur.582..497P} is underestimated, %by $\approx 10\,\%$, 
and/or  (3) AU Mic is differentially rotating.  
%It is not easy to say which scenario is more likely at this point, and further observations of AU Mic would provide some key information on the above possibilities. 
One may be able to constrain the presence of differential rotation by
continuous photometric monitoring of AU Mic.

%To ensure the robustness of our conclusion that the system has a low obliquity, we repeated the Doppler shadow analysis imposing a prior on $v\sin i$ based on the photometric rotation period (i.e., $v\sin i=7.81\pm0.31$).  After the MCMC analysis for the observed residual CCF map, we found that result
%was unchanged, reproducing a good spin-orbit alignment ($\lambda=1.3_{-4.7}^{+5.6}$ degrees). 

\subsection{He I line modeling}
\label{sec:hei}

\begin{figure*}[t]
\centering
\includegraphics[width=13.5 cm]{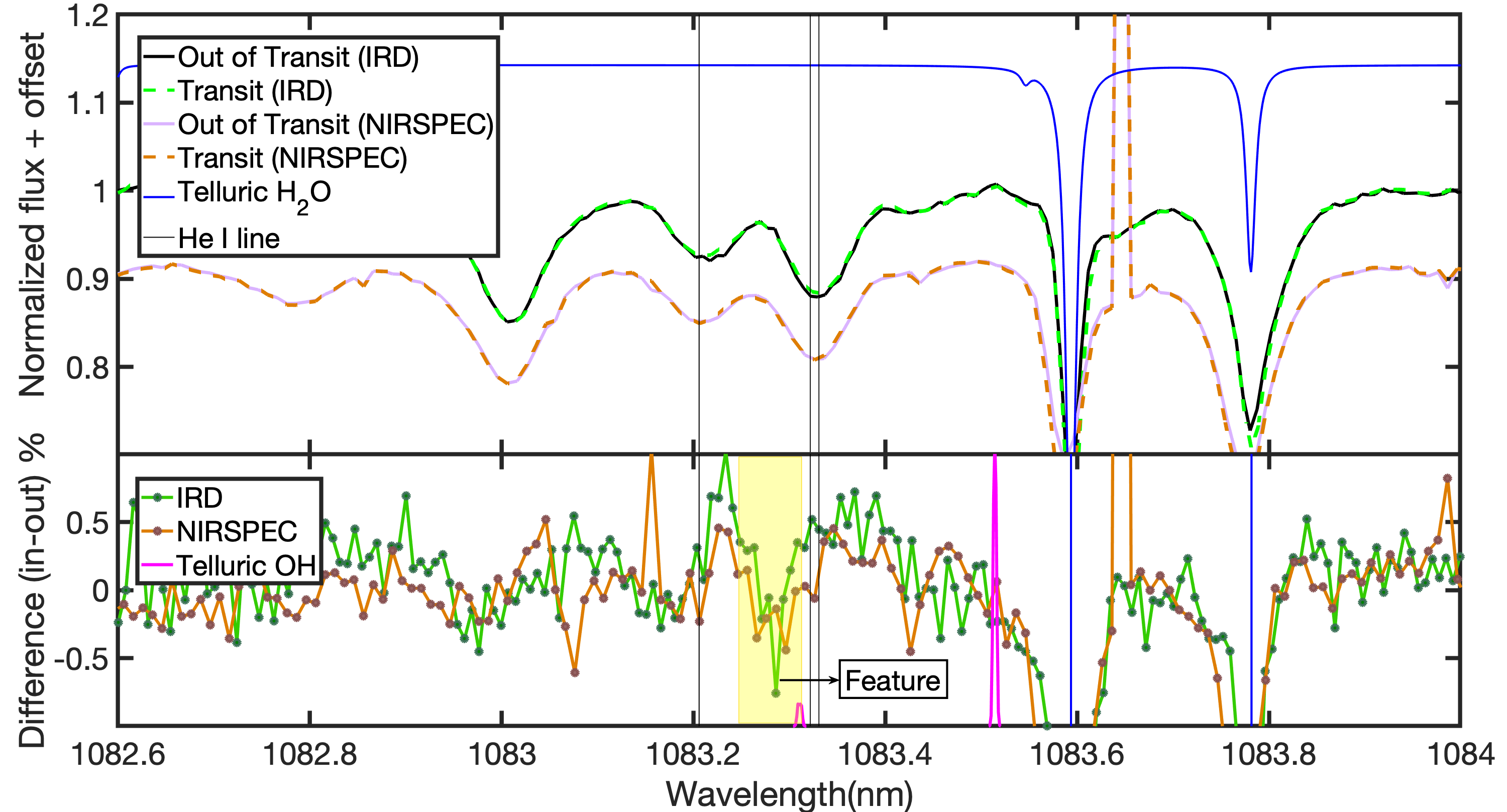}
\caption{{\bf Top}: Spectra of AU Mic in the vicinity of the \hei\ lines (vertical black lines) obtained in and out of the transit of ``b" by IRD (green and black) and NIRSPEC (orange and violet).  Predicted telluric H$_2$O (blue) and OH (magenta) lines are from \citet{Breckinridge1973} and \citet{Noll2012,Jones2013}, respectively.  {\bf Bottom}: The corresponding difference spectra (in-out) in percentage. The yellow-shaded region shows the $\sim0.1$\,nm-wide feature.
}
\label{fig:he_diff}
\end{figure*}

The unresolved $J=1,2$ doublet and the bluer $J=0$ line of neutral orthohelium are both detected (Figure \ref{fig:he_diff}).  The relative equivalent widths (EWs) suggest that the doublet is optically thick and that the absorption arises from a relatively small filling factor on the star \citep{Andretta2017}.  The total EW (107 m\AA{}) is small for the star's X-ray activity \citep[$L_X/L_{\rm bol}= -2.82$;][]{Plavchan2009}, possibly because emission partially fills in the line \citep{Smith2016}.   

We used the transit ephemeris and duration from \citet{2020Natur.582..497P} to sum subsets of spectra inside and outside of the transit.  Individual IRD spectra were adjusted to account for the varying Doppler shift of the planet ($\pm 5$\,km\,s$^{-1}$). Due to the lower resolution, this correction was unnecessary for the NIRSPEC spectra.  A comparison of the summed spectra taken during transit and post-egress indicate no transit-related absorption at the wavelength of the strong doublet (Figure \ref{fig:he_diff}). To limit the presence of any planet-associated absorption, we assumed a Gaussian profile with thermal broadening of 1000-30000~K for the IRD data. We adopted an RMS value of 1.2\% of the IRD difference spectrum for a 10 \AA{} region in the vicinity of the line. A similar model was implemented on NIRSPEC data. Using a $\chi^2$ analysis we limit the EW of any planet-associated signal to 4.4 and 3.7 m\AA{} (99\% confidence) from the IRD and NIRSPEC spectra, respectively.

A $\sim0.1$\,nm-wide feature about 0.05\,nm blue-ward of the doublet center is apparent in both the IRD and NIRSPEC spectra (Figure \ref{fig:he_diff}).  Since it appears in both independent data sets, it is not an artifact of the instrument or reduction routine.  Such a feature could be Doppler-shifted absorption in a gaseous tail moving relative to the star at $\approx$-15\,km\,s$^{-1}$, but it could also be a systematic, i.e.: (a) telluric variation over the transit/airmass; (b) a residual from changing Doppler shift of the stellar He\,I line over a few hours; (c) planetary occultation of active regions of the star responsible for the He\,I signal; (d) changing of the stellar line shape due to active regions appearing or disappearing from the stellar limb; and (e) changes in the stellar line due to activity.  We consider (a) unlikely; there are no known H$_2$O lines in the vicinity (Figure \ref{fig:he_diff}). A weak OH line coincides with the feature but should have been subtracted by the rapid beam switching during the NIRSPEC observations and other, stronger OH lines do not produce artifacts.  We rule out (b) by simulating the effect, i.e., dividing the out-of-transit spectrum by a second that is Doppler shifted by the barycenter change of 440 m s$^{-1}$ over the observation interval and finding negligible difference.   We rule out (c) because we see no significant variation in the in-transit signal when binned into four intervals during the transit.  We consider (d) to be unlikely because the rotation period (4.86\,days) is long compared to the observation interval.  

\begin{figure}
\centering
\includegraphics[width=8.6 cm]{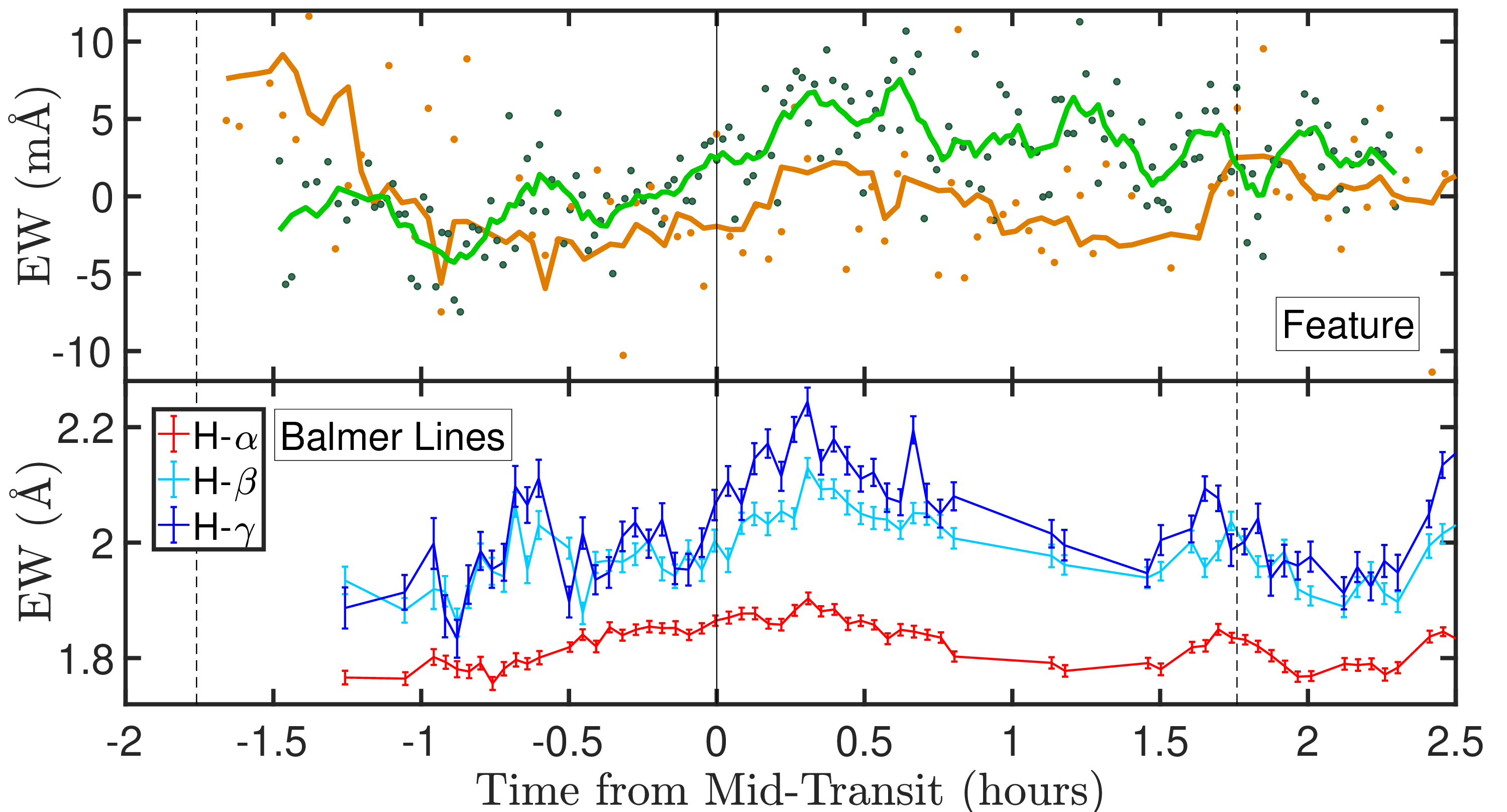}
\caption{{\bf Top}: Strength of the feature between 1083.225 and 1083.325\,nm vs. time as measured in IRD (green) and NIRSPEC (orange) spectra.   The solid lines represent 7-point, first-order Savitzky-Golay filtered versions. {\bf Bottom}: EW of hydrogen Balmer lines: H-$\alpha$ (red), H-$\beta$ (aqua) and H-$\gamma$ (blue) as measured in SNIFS spectra.  In both panels a positive value represents increasing \emph{emission}.
}
\label{fig:ew_feature}
\end{figure}

Explanation (e) remains viable because our Balmer line series measurements with SNIFS show multiple rapid rises and slower (hour-timescale) falls (Figure \ref{fig:ew_feature}) due to variable activity including numerous small flares \citep{Robinson2001}.   Flaring could produce \hei\ emission which is blue-shifted due to plasma motion (15 km\,s$^{-1}$ is twice the rotational velocity of the star).  The strength of the feature in IRD spectra is significantly correlated with the H$\alpha$ EW from SNIFS spectra (Spearman rank $\rho = 0.24$ and $p = 1.9 \times 10^{-3}$; Pearson's $R = 0.26$ with $p=5.0\times 10^{-4}$), supporting this scenario \citep[see, e.g. ][]{Guilluy2020}.  NIRSPEC derived values are not correlated (Spearman rank $\rho = -0.03$ and $p = 0.80$; Pearson's $R = 0.07$ with $p=0.53$), possibly due to lower instrument resolution and an early decline seen in the NIRSPEC but not the IRD data (Figure \ref{fig:ew_feature}).  

%%%%%%%%%%%%%%%%%%%%%%%%%%
%%%%%%%%%%%%%%%%%%%%%%%%%%

\begin{figure}
\centering
\includegraphics[width=8.6cm]{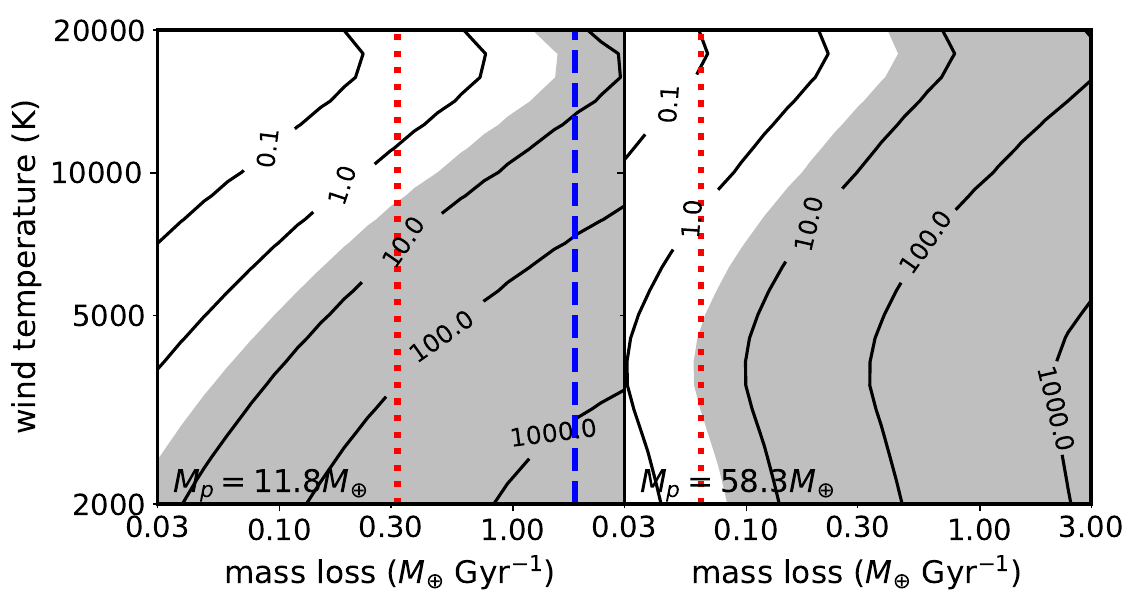}
\caption{Contours of predicted EW in m\AA{} of the 1083 nm \hei\ triplet vs. mass loss rate and wind temperature for an isothermal, solar-composition model of an escaping atmosphere from AU Mic b, for the current best-fit value (left) and a 99\% upper limit on planet mass based on RV data (P. Plavchan, private comm.).  The grey region is ruled out by our transit observations.  The red dotted lines are the mass loss rates for energy-limited escape, assuming 10\% efficiency \citep{Shematovich2014} and the blue dashed line is the prediction based on the hydrodynamic models of \citet{Kubyshkina2018}; the estimate in the right panel is off scale to the left.
}
\label{fig:model}
\end{figure}

The planetary \hei\ line was modeled with an isothermal Parker wind with a solar-like H/He ratio (10.5) as described in \citet{Gaidos2020} \citep[see also][]{Oklopovcic2018}.  The abundance of triplet \hei\ is governed by photoionization of ground-state \hei\ by EUV photons and photoionization of the triplet \hei\ by FUV and NUV photons.  No complete UV spectrum of AU Mic is available, so we constructed one by combining an \emph{EUVE} spectrum \citep{Monsignori1996} with a synthetic spectrum of the M1.5-type dwarf GJ 832 \citep{Fontenla2016} and adjusting the EUV (90-360\,\AA), Ly $\alpha$ (1216\AA{}), FUV (1340-1800\AA) and NUV (1700-3000\AA) domains by an \emph{EUVE} observation \citep{France2018}, \emph{HST} observations \citep{Linsky2014}, and \emph{GALEX} observations analyzed by \citet{Schneider2018}.  The spectrum was uniformly adjusted within these wavelength ranges to match the observed fluxes.  Both the sonic radius (which sets the scale of the flow) and Roche radius (beyond which it is assumed the flow is dispersed) depend on the planet mass, and through them the \hei\ signal at a given mass loss rate is mass-dependent.  The mass of ``b" is not precisely determined and we considered the best-fit value to available RV data (11.8$M_{\oplus}$) and a 99\% confidence upper limit (57.3$M_{\oplus}$; P. Plavchan, in prep.).  Figure \ref{fig:model} plots contours of constant EW for the two planet mass cases, showing that mass loss rates of $>0.15-0.45$ \mearth\,Gyr$^{-1}$ are ruled out for a wind temperature of 10000\,K, with (cooler/hotter) winds (more/less) restricted in terms of mass loss rate.

%%%%%%%%%%%%%%%%%%%%%%%%%%

\section{Discussion and Summary} \label{sec:discussion}

At $\approx22$\,Myr, the AU Mic system should not have undergone tidal spin-orbit
realignment; the tidal realignment times of a Neptune-mass planet in a $P=8.46$ 
day orbit is longer than a few Gyr \citep[e.g.,][]{2010ApJ...718L.145W, 2012ApJ...757...18A}, and the system should retain any primordial obliquity.  Therefore, the inferred low stellar obliquity suggests that the system has not experienced an event such as planet-planet scatterings \citep[e.g.,][]{2008ApJ...686..580C, 2011ApJ...742...72N}, and that AU Mic b likely formed {\it in situ} or migrated to its present location by torques from a primordial disk \citep[e.g.,][]{1996Natur.380..606L, 2012ARA&A..50..211K}. AU Mic is one of the youngest systems known to host a transiting close-in planet, making our measurements an important benchmark to compare to models of the dynamical evolution of close-in planets. 

One caveat of our obliquity analysis is that the residual CCF map is affected by activity-induced variations, which we did not account for in deriving $\lambda$. AU Mic is an active, spotted star, and CCF features will evolve with spots. In addition, small flares can distort the CCF profile, and add systematic errors. Fortunately, those effects are known to be mitigated to be a factor $2-4$ smaller in the infrared compared with visible due to reduced contrasts of surface spots and faculae \citep[e.g.,][]{2012ApJ...761..164C, 2019RNAAS...3...89B}.   Indeed, Figure  \ref{fig:DT} does not exhibit evident spot- and flare-like features in the residual line profile. The reason for modulation seen in the $H-$band CCF map is not known, but we suspect it is caused by imperfect processing of telluric lines and/or detector's persistence, both of which are more significant in the $H-$band.  Observations of additional transits are needed to evaluate these effects.

We detect no He\,I absorption during the transit of AU Mic b that could be associated with its atmosphere.  A blue-shifted feature is unlikely to be an artifact of our instruments, or tellurics, but appears to be the product of stellar activity. This highlights the challenge of planet transit spectroscopy for young, active stellar hosts \citep{Cauley2018} and the imperative of obtaining simultaneous data on activity \citep{Guilluy2020}.  Our limits on the atmosphere escape rate from the non-detection of He\,I could challenge models which explain a ``Neptune desert" by XUV photoevaporative escape, primarily during the first 100\,Myr when the host star is most active \citep{Owen2019}.  Loss of the $\sim$1 \mearth{} of H responsible for the large radii of Neptune-like planets in this time would require $10$ \mearth{} Gyr$^{-1}$.  AU Mic b orbits outside the ``desert", nevertheless substantial loss would be expected. We estimated escape rates using a combined X-ray, EUV (90-360\,\AA) and Lyman-$\alpha$ irradiances (46.5 W m$^{-2}$) and (a) the energy-limited escape relation of \citet{Erkaev2007} with an efficiency $\eta = 10$\%  \citep{Shematovich2014}, and (b) the hydrodynamic model-based relations of \citet{Kubyshkina2018}.  For the best-estimate planet mass (11.8\,\mearth{}) we can rule out both estimates (0.31 and 1.85\,\mearth\,Gyr$^{-1}$, respectively) for most wind temperatures considered (left panel of Fig. \ref{fig:model}).  However, in the case of a planet mass equal to the 99\% confidence upper limit (58.3\,\mearth, right panel of Fig. \ref{fig:model}), the predicted mass loss rate would be only 0.064 and $5 \times 10^{-3}$\,\mearth\,Gyr$^{-1}$, respectively) and our observations are not constraining.  Refinement of the planet mass and an accounting of the effects of a non-spherical geometry, interaction with the stellar wind and radiation pressure are needed. Unlike the case for K2-100b \citep{Gaidos2020}, our model does not predict complete photoionization of H in the wind; rather for escape rates near the upper limit resonant scattering by H I should produce broad ($\sim1$\AA) absorption in the Lyman $\alpha$ line during a transit and could be detected by \emph{HST}.

%Through the simultaneous observations of the spectroscopic transit for AU Mic b 
%using three ground-based facilities, we investigated the stellar obliquity, 
%excess absorption by the planet, and activity indicators of the host star, 
%putting tight constraints on those quantities. 
%Since AU Mic is one of the youngest systems hosting a transiting close-in planet, 
%our measurements would become an important benchmark to discuss the dynamical evolution
%and star-planet interactions, including atmospheric loss, of close-in planets. 
%Given that AU Mic b is also one of the closest exoplanets known to date, 
%further observations such as search for outer companions and transmission spectroscopy 
%from the space are encouraged, which would unveil the overall picture of this 
%remarkable system. 

\acknowledgments

This work was supported by JSPS KAKENHI Grant Numbers JP19K14783, JP18H05442, JP15H02063, JP22000005, JP18H01265, and JP18H05439, JST PRESTO Grant Number JPMJPR1775, and by the Astrobiology Center Program of National Institutes of Natural Sciences (NINS) (Grant Number AB311017). E.G. carried out some of this research while a participant in the ``ExoStar" program at the Kavli Institute for Theoretical Physics, which is supported in part by the National Science Foundation under Grant No. NSF PHY-1748958.  E.G. thanks the UC Santa Barbara Department of Physics, especially Glenn Schiferl, for their hospitality and assistance in conducting remote observations with Keck-2, and fellow observer Erik Petigura for obtaining darks and flat fields.  The W. M. Keck Observatory is operated as a scientific partnership among the California Institute of Technology, the University of California and the National Aeronautics and Space Administration. The Observatory was made possible by the generous financial support of the W. M. Keck Foundation.  SNIFS on the UH 2.2-m telescope is part of the Nearby Supernova Factory project, a scientific collaboration among the Centre de Recherche Astronomique de Lyon, Institut de Physique Nucl\'{e}aire de Lyon, Laboratoire de Physique Nucl\'{e}aire et des Hautes Energies, Lawrence Berkeley National Laboratory, Yale University, University of Bonn, Max Planck Institute for Astrophysics, Tsinghua Center for Astrophysics, and the Centre de Physique des Particules de Marseille.  The data analysis was carried out, in part, on the Multi-wavelength Data Analysis System operated by the Astronomy Data Center (ADC), National Astronomical Observatory of Japan. This research made use of {\tt Astropy}\footnote{http://www.astropy.org} a community-developed core {\it Python} package for Astronomy \citep{astropy:2018}, and provided by the High Energy Astrophysics Science Archive Research Center (HEASARC), which is a service of the Astrophysics Science Division at NASA/GSFC.

%% To help institutions obtain information on the effectiveness of their 
%% telescopes the AAS Journals has created a group of keywords for telescope 
%% facilities.
%
%% Following the acknowledgments section, use the following syntax and the
%% \facility{} or \facilities{} macros to list the keywords of facilities used 
%% in the research for the paper.  Each keyword is check against the master 
%% list during copy editing.  Individual instruments can be provided in 
%% parentheses, after the keyword, but they are not verified.

\vspace{5mm}
\facilities{Subaru(IRD), Keck II(NIRSPEC)}

%% Similar to \facility{}, there is the optional \software command to allow 
%% authors a place to specify which programs were used during the creation of 
%% the manuscript. Authors should list each code and include either a
%% citation or url to the code inside ()s when available.

\software{
{\tt IRAF} \citep{1993ASPC...52..173T}
}

%% Appendix material should be preceded with a single \appendix command.
%% There should be a \section command for each appendix. Mark appendix
%% subsections with the same markup you use in the main body of the paper.

%% Each Appendix (indicated with \section) will be lettered A, B, C, etc.
%% The equation counter will reset when it encounters the \appendix
%% command and will number appendix equations (A1), (A2), etc. The
%% Figure and Table counter will not reset.

\clearpage

\appendix

\section{Analysis of the Effective RVs} \label{sec:app1}

\begin{figure}
\centering
\includegraphics[width=8.5cm]{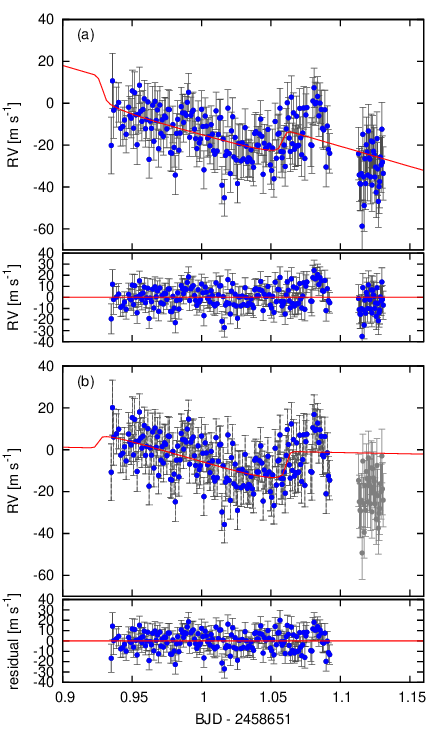}
\caption{
Results of the RV analysis and modeling of the effective RVs for the RM effect. 
Panel (a) presents the result of the RM fit using all the RV data during the
transit night, while panel (b) shows the fitting result for the RV data set
before observing the telluric standard star. In both fits, a Gaussian prior
is imposed on $v\sin i$. 
}
\label{fig:RV}
\end{figure}

RVs for AU Mic, derived in Section \ref{sec:RV}, are plotted in Figure \ref{fig:RV}.  The typical internal RV error during the night was $8-10$ m s$^{-1}$. In order to constrain the projected stellar obliquity $\lambda$, we fitted the observed RVs using the analytic RM model by \citet{2011ApJ...742...69H}. Since the obliquity measurement by fitting the effective RVs sensitively depends on the RV baseline, we allowed the RV semi-amplitude $K$ by the Keplerian planetary orbit to float freely, assuming a circular orbit. Although \citet{2020Natur.582..497P} reported a constraint on AU Mic b's mass from the RV monitoring, we do not use any prior information on the RV variation due to the lack of knowledge on the instantaneous RV variations by stellar activity, especially for the wavelength range covered by IRD. 

%%%%%%%%%%%%%%%%%%%
%%%%%%%%%%%%%%%%%%%
We first attempted to fit all the observed RVs during the night without a prior on 
$v\sin i$. Due to the lack of the transit ingress and RV baseline, 
however, the fit did not 
converge, exhibiting a strong correlation among $K$, $v\sin i$, and $\lambda$. We thus 
decided to impose a prior on $v\sin i$ based on spectroscopy. 
Following \citet{2020ApJ...890L..27H}, we analyzed the mean out-of-transit  
CCF of the IRD spectra for AU Mic, and derived $v\sin i$ by comparing the observed
CCF with a number of theoretically simulated CCFs for different $v\sin i$. 
The fitting result suggests $v\sin i=8.38_{-0.17}^{+0.19}$ km s$^{-1}$, which is compatible
with the value reported in the discovery paper ($v\sin i=8.7\pm 0.2$ km s$^{-1}$). To take
into account the systematic error in the derived $v\sin i$, especially arising from the
uncertainty in the stellar macroturbulent velocity, we enlarged the error bar for $v\sin i$
and employed the Gaussian prior of $v\sin i=8.4\pm 0.4$ km s$^{-1}$ in the RM fit. 
In fitting the observed RVs by the MCMC analysis, %\citep{2016ApJ...825...53H}, 
we also imposed Gaussian priors on the mid-transit time
($T_c=2458651.99351\pm0.00070$), the scaled semi-major axis ($a/Rs=19.1\pm1.7$),
and the transit impact parameters ($b=0.16\pm0.14$) based on the values in 
\citet{2020Natur.582..497P}. 
The optimized parameters in the MCMC analysis are $K$, $v\sin i$, $\lambda$, 
$T_c$, $a/R_s$, $b$, and RV offset $\gamma$ for the IRD data set.

The red solid line in panel (a) of Figure \ref{fig:RV} draws the best-fit model for the
observed RVs. The model yielded $\lambda=91\pm 5$ degress, suggesting
a significant spin-orbit misalignment of the system. However, the model also implied
$K=260_{-36}^{+33}$ m s$^{-1}$, which is inconsistent with what \citet{2020Natur.582..497P} found. 
This unusually large $K$ made us wonder that our analysis of the observed RVs could be 
affected by the systematic effects associated with the lack of baseline data. 
Specifically, the RV data points beyond 
$\mathrm{BJD}- 2458651=1.1$ are mostly responsible for the large $K$, but 
those IRD spectra were taken with very high airmass at the very end of the night during 
twilight. We thus performed the same analysis above only using RV data before 
$\mathrm{BJD}- 2458651=1.1$, and derived the obliquity $\lambda$ along with the
other parameters. The MCMC fit for this case resulted in $K=17_{-73}^{+75}$ m s$^{-1}$
and $\lambda=69_{-8}^{+7}$ degrees. Its best-fit model is shown by the red solid line 
in panel (b) of Figure \ref{fig:RV}.

Lastly, we repeated the MCMC analysis for the limited set of RV data 
($\mathrm{BJD}- 2458651<1.1$) 
without a prior on $v\sin i$. Although the posterior distribution still shows a 
degeneracy between $v\sin i$ and $\lambda$, a relatively small value was favored for $K$. 
We obtained $\lambda=-45_{-81}^{+114}$ degrees and $v\sin i=7.5_{-2.8}^{+17.4}$
km s$^{-1}$ from the MCMC fit.  
The results of all these analyses suggest that the stellar obliquity estimated from our RV 
data sensitively depends on both $v\sin i$ and the overall RV baseline 
during the night, and it is not straightforward to gain a reliable (accurate) estimate 
for $\lambda$, mostly owing to the lack of baseline RV data. 
Therefore, we decided not to use the RV data and 
resort to an alternative approach (Doppler-shadow analysis) for
the measurement of $\lambda$. 

%% For this sample we use BibTeX plus aasjournals.bst to generate the
%% the bibliography. The sample63.bib file was populated from ADS. To
%% get the citations to show in the compiled file do the following:
%%
%% pdflatex sample63.tex
%% bibtext sample63
%% pdflatex sample63.tex
%% pdflatex sample63.tex

%\bibliography{hirano2020_reference,other}{}

\begin{thebibliography}{}
\expandafter\ifx\csname natexlab\endcsname\relax\def\natexlab#1{#1}\fi
\providecommand{\url}[1]{\href{#1}{#1}}
\providecommand{\dodoi}[1]{doi:~\href{http://doi.org/#1}{\nolinkurl{#1}}}
\providecommand{\doeprint}[1]{\href{http://ascl.net/#1}{\nolinkurl{http://ascl%
.net/#1}}}
\providecommand{\doarXiv}[1]{\href{https://arxiv.org/abs/#1}{\nolinkurl{https:%
//arxiv.org/abs/#1}}}

\bibitem[{{Albrecht} {et~al.}(2012){Albrecht}, {Winn}, {Johnson}, {Howard},
  {Marcy}, {Butler}, {Arriagada}, {Crane}, {Shectman}, {Thompson}, {Hirano},
  {Bakos}, \& {Hartman}}]{2012ApJ...757...18A}
{Albrecht}, S., {Winn}, J.~N., {Johnson}, J.~A., {et~al.} 2012, \apj, 757, 18,
  \dodoi{10.1088/0004-637X/757/1/18}

\bibitem[{{Andretta} {et~al.}(2017){Andretta}, {Giampapa}, {Covino}, {Reiners},
  \& {Beeck}}]{Andretta2017}
{Andretta}, V., {Giampapa}, M.~S., {Covino}, E., {Reiners}, A., \& {Beeck}, B.
  2017, \apj, 839, 97, \dodoi{10.3847/1538-4357/aa6a14}

\bibitem[{{Beichman} {et~al.}(2019){Beichman}, {Hirano}, {David}, {Kotani},
  {Hillenbrand}, {Vasisht}, {Ciardi}, {Harakawa}, {Kudo}, {Omiya}, {Kuzuhara},
  \& {Tamura}}]{2019RNAAS...3...89B}
{Beichman}, C., {Hirano}, T., {David}, T.~J., {et~al.} 2019, Research Notes of
  the American Astronomical Society, 3, 89, \dodoi{10.3847/2515-5172/ab2c9d}

\bibitem[{{Breckinridge} \& {Hall}(1973)}]{Breckinridge1973}
{Breckinridge}, J.~B., \& {Hall}, D. N.~B. 1973, \solphys, 28, 15,
  \dodoi{10.1007/BF00152906}

\bibitem[{{Cauley} {et~al.}(2018){Cauley}, {Kuckein}, {Redfield}, {Shkolnik},
  {Denker}, {Llama}, \& {Verma}}]{Cauley2018}
{Cauley}, P.~W., {Kuckein}, C., {Redfield}, S., {et~al.} 2018, \aj, 156, 189,
  \dodoi{10.3847/1538-3881/aaddf9}

\bibitem[{{Chatterjee} {et~al.}(2008){Chatterjee}, {Ford}, {Matsumura}, \&
  {Rasio}}]{2008ApJ...686..580C}
{Chatterjee}, S., {Ford}, E.~B., {Matsumura}, S., \& {Rasio}, F.~A. 2008, \apj,
  686, 580, \dodoi{10.1086/590227}

\bibitem[{{Collier Cameron} {et~al.}(2010){Collier Cameron}, {Guenther},
  {Smalley}, {McDonald}, {Hebb}, {Andersen}, {Augusteijn}, {Barros}, {Brown},
  {Cochran}, {Endl}, {Fossey}, {Hartmann}, {Maxted}, {Pollacco}, {Skillen},
  {Telting}, {Waldmann}, \& {West}}]{2010MNRAS.407..507C}
{Collier Cameron}, A., {Guenther}, E., {Smalley}, B., {et~al.} 2010, \mnras,
  407, 507, \dodoi{10.1111/j.1365-2966.2010.16922.x}

\bibitem[{{Crockett} {et~al.}(2012){Crockett}, {Mahmud}, {Prato},
  {Johns-Krull}, {Jaffe}, {Hartigan}, \& {Beichman}}]{2012ApJ...761..164C}
{Crockett}, C.~J., {Mahmud}, N.~I., {Prato}, L., {et~al.} 2012, \apj, 761, 164,
  \dodoi{10.1088/0004-637X/761/2/164}

\bibitem[{{David} {et~al.}(2019){David}, {Petigura}, {Luger}, {Foreman-Mackey},
  {Livingston}, {Mamajek}, \& {Hillenbrand}}]{David2019}
{David}, T.~J., {Petigura}, E.~A., {Luger}, R., {et~al.} 2019, \apjl, 885, L12,
  \dodoi{10.3847/2041-8213/ab4c99}

\bibitem[{{David} {et~al.}(2016){David}, {Hillenbrand}, {Petigura},
  {Carpenter}, {Crossfield}, {Hinkley}, {Ciardi}, {Howard}, {Isaacson}, {Cody},
  {Schlieder}, {Beichman}, \& {Barenfeld}}]{David2016}
{David}, T.~J., {Hillenbrand}, L.~A., {Petigura}, E.~A., {et~al.} 2016, \nat,
  534, 658, \dodoi{10.1038/nature18293}

\bibitem[{{Erkaev} {et~al.}(2007){Erkaev}, {Kulikov}, {Lammer}, {Selsis},
  {Langmayr}, {Jaritz}, \& {Biernat}}]{Erkaev2007}
{Erkaev}, N.~V., {Kulikov}, Y.~N., {Lammer}, H., {et~al.} 2007, \aap, 472, 329,
  \dodoi{10.1051/0004-6361:20066929}

\bibitem[{{Fabrycky} \& {Tremaine}(2007)}]{2007ApJ...669.1298F}
{Fabrycky}, D., \& {Tremaine}, S. 2007, \apj, 669, 1298, \dodoi{10.1086/521702}

\bibitem[{{Fontenla} {et~al.}(2016){Fontenla}, {Linsky}, {Witbrod}, {France},
  {Buccino}, {Mauas}, {Vieytes}, \& {Walkowicz}}]{Fontenla2016}
{Fontenla}, J.~M., {Linsky}, J.~L., {Witbrod}, J., {et~al.} 2016, \apj, 830,
  154, \dodoi{10.3847/0004-637X/830/2/154}

\bibitem[{{France} {et~al.}(2018){France}, {Arulanantham}, {Fossati}, {Lanza},
  {Loyd}, {Redfield}, \& {Schneider}}]{France2018}
{France}, K., {Arulanantham}, N., {Fossati}, L., {et~al.} 2018, \apjs, 239, 16,
  \dodoi{10.3847/1538-4365/aae1a3}

\bibitem[{{Fulton} {et~al.}(2017){Fulton}, {Petigura}, {Howard}, {Isaacson},
  {Marcy}, {Cargile}, {Hebb}, {Weiss}, {Johnson}, {Morton}, {Sinukoff},
  {Crossfield}, \& {Hirsch}}]{2017AJ....154..109F}
{Fulton}, B.~J., {Petigura}, E.~A., {Howard}, A.~W., {et~al.} 2017, \aj, 154,
  109, \dodoi{10.3847/1538-3881/aa80eb}

\bibitem[{{Gaidos} {et~al.}(2020){Gaidos}, {Hirano}, {Mann}, {Owens}, {Berger},
  {France}, {Vanderburg}, {Harakawa}, {Hodapp}, {Ishizuka}, {Jacobson},
  {Konishi}, {Kotani}, {Kudo}, {Kurokawa}, {Kuzuhara}, {Nishikawa}, {Omiya},
  {Serizawa}, {Tamura}, \& {Ueda}}]{Gaidos2020}
{Gaidos}, E., {Hirano}, T., {Mann}, A.~W., {et~al.} 2020, \mnras,
  \dodoi{10.1093/mnras/staa918}

\bibitem[{{Ginzburg} {et~al.}(2018){Ginzburg}, {Schlichting}, \&
  {Sari}}]{Ginzburg2018}
{Ginzburg}, S., {Schlichting}, H.~E., \& {Sari}, R. 2018, \mnras, 476, 759,
  \dodoi{10.1093/mnras/sty290}

\bibitem[{{Grady} {et~al.}(2020){Grady}, {Wisniewski}, {Schneider},
  {Boccaletti}, {Gaspar}, {Debes}, {Hines}, {Stark}, {Thalmann}, {Lagrange},
  {Augereau}, {Sezestre}, {Milli}, {Henning}, \& {Kuchner}}]{Grady2020}
{Grady}, C.~A., {Wisniewski}, J.~P., {Schneider}, G., {et~al.} 2020, \apjl,
  889, L21, \dodoi{10.3847/2041-8213/ab65bb}

\bibitem[{{Guilluy} {et~al.}(2020){Guilluy}, {Andretta}, {Borsa}, {Giacobbe},
  {Sozzetti}, {Covino}, {Bourrier}, {Fossati}, {Bonomo}, {Esposito},
  {Giampapa}, {Harutyunyan}, {Rainer}, {Brogi}, {Bruno}, {Claudi}, {Frustagli},
  {Lanza}, {Mancini}, {Pino}, {Poretti}, {Scandariato}, {Affer}, {Baffa},
  {Baruffolo}, {Benatti}, {Biazzo}, {Bignamini}, {Boschin}, {Carleo},
  {Cecconi}, {Cosentino}, {Damasso}, {Desidera}, {Falcini}, {Martinez
  Fiorenzano}, {Ghedina}, {Gonz{\'a}lez-{\'A}lvarez}, {Guerra}, {Hernandez},
  {Leto}, {Maggio}, {Malavolta}, {Maldonado}, {Micela}, {Molinari},
  {Nascimbeni}, {Pagano}, {Pedani}, {Piotto}, \& {Reiners}}]{Guilluy2020}
{Guilluy}, G., {Andretta}, V., {Borsa}, F., {et~al.} 2020, arXiv e-prints,
  arXiv:2005.05676.
\newblock \doarXiv{2005.05676}

\bibitem[{{Hirano} {et~al.}(2010){Hirano}, {Suto}, {Taruya}, {Narita}, {Sato},
  {Johnson}, \& {Winn}}]{2010ApJ...709..458H}
{Hirano}, T., {Suto}, Y., {Taruya}, A., {et~al.} 2010, \apj, 709, 458,
  \dodoi{10.1088/0004-637X/709/1/458}

\bibitem[{{Hirano} {et~al.}(2011){Hirano}, {Suto}, {Winn}, {Taruya}, {Narita},
  {Albrecht}, \& {Sato}}]{2011ApJ...742...69H}
{Hirano}, T., {Suto}, Y., {Winn}, J.~N., {et~al.} 2011, \apj, 742, 69,
  \dodoi{10.1088/0004-637X/742/2/69}

\bibitem[{{Hirano} {et~al.}(2020){Hirano}, {Gaidos}, {Winn}, {Dai}, {Fukui},
  {Kuzuhara}, {Kotani}, {Tamura}, {Hjorth}, {Albrecht}, {Huber}, {Bolmont},
  {Harakawa}, {Hodapp}, {Ishizuka}, {Jacobson}, {Konishi}, {Kudo}, {Kurokawa},
  {Nishikawa}, {Omiya}, {Serizawa}, {Ueda}, \& {Weiss}}]{2020ApJ...890L..27H}
{Hirano}, T., {Gaidos}, E., {Winn}, J.~N., {et~al.} 2020, \apjl, 890, L27,
  \dodoi{10.3847/2041-8213/ab74dc}

\bibitem[{{Johnson} {et~al.}(2017){Johnson}, {Cochran}, {Addison}, {Tinney}, \&
  {Wright}}]{2017AJ....154..137J}
{Johnson}, M.~C., {Cochran}, W.~D., {Addison}, B.~C., {Tinney}, C.~G., \&
  {Wright}, D.~J. 2017, \aj, 154, 137, \dodoi{10.3847/1538-3881/aa8462}

\bibitem[{{Jones} {et~al.}(2013){Jones}, {Noll}, {Kausch}, {Szyszka}, \&
  {Kimeswenger}}]{Jones2013}
{Jones}, A., {Noll}, S., {Kausch}, W., {Szyszka}, C., \& {Kimeswenger}, S.
  2013, \aap, 560, A91, \dodoi{10.1051/0004-6361/201322433}

\bibitem[{{Kley} \& {Nelson}(2012)}]{2012ARA&A..50..211K}
{Kley}, W., \& {Nelson}, R.~P. 2012, \araa, 50, 211,
  \dodoi{10.1146/annurev-astro-081811-125523}

\bibitem[{{Kotani} {et~al.}(2018){Kotani}, {Tamura}, {Nishikawa}, {Ueda},
  {Kuzuhara}, {Omiya}, {Hashimoto}, {Ishizuka}, {Hirano}, {Suto}, {Kurokawa},
  {Kokubo}, {Mori}, {Tanaka}, {Kashiwagi}, {Konishi}, {Kudo}, {Sato},
  {Jacobson}, {Hodapp}, {Hall}, {Aoki}, {Usuda}, {Nishiyama}, {Nakajima},
  {Ikeda}, {Yamamuro}, {Morino}, {Baba}, {Hosokawa}, {Ishikawa}, {Narita},
  {Kokubo}, {Hayano}, {Izumiura}, {Kambe}, {Kusakabe}, {Kwon}, {Ikoma}, {Hori},
  {Genda}, {Fukui}, {Fujii}, {Kawahara}, {Olivier}, {Jovanovic}, {Harakawa},
  {Hayashi}, {Hidai}, {Machida}, {Matsuo}, {Nagata}, {Ogihara}, {Takami},
  {Takato}, {Terada}, \& {Oh}}]{2018SPIE10702E..11K}
{Kotani}, T., {Tamura}, M., {Nishikawa}, J., {et~al.} 2018, in \procspie, Vol.
  10702, Ground-based and Airborne Instrumentation for Astronomy VII, 1070211,
  \dodoi{10.1117/12.2311836}

\bibitem[{{Kubyshkina} {et~al.}(2018){Kubyshkina}, {Fossati}, {Erkaev},
  {Cubillos}, {Johnstone}, {Kislyakova}, {Lammer}, {Lendl}, \&
  {Odert}}]{Kubyshkina2018}
{Kubyshkina}, D., {Fossati}, L., {Erkaev}, N.~V., {et~al.} 2018, \apjl, 866,
  L18, \dodoi{10.3847/2041-8213/aae586}

\bibitem[{{Kuzuhara} {et~al.}(2018){Kuzuhara}, {Hirano}, {Kotani}, {Ishizuka},
  {Omiya}, {Konishi}, {Kudo}, {Nishikawa}, {Ueda}, {Hosokawa}, {Kusakabe},
  {Kurokawa}, {Kokubo}, {Mori}, {Tanaka}, {Jacobson}, {Hodapp}, \&
  {Tamura}}]{2018SPIE10702E..60K}
{Kuzuhara}, M., {Hirano}, T., {Kotani}, T., {et~al.} 2018, in \procspie, Vol.
  10702, Ground-based and Airborne Instrumentation for Astronomy VII, 1070260,
  \dodoi{10.1117/12.2311832}

\bibitem[{{Lantz} {et~al.}(2004){Lantz}, {Aldering}, {Antilogus}, {Bonnaud},
  {Capoani}, {Castera}, {Copin}, {Dubet}, {Gangler}, {Henault}, {Lemonnier},
  {Pain}, {Pecontal}, {Pecontal}, \& {Smadja}}]{Lantz2004}
{Lantz}, B., {Aldering}, G., {Antilogus}, P., {et~al.} 2004, in \procspie, Vol.
  5249, Optical Design and Engineering, ed. L.~{Mazuray}, P.~J. {Rogers}, \&
  R.~{Wartmann}, 146--155, \dodoi{10.1117/12.512493}

\bibitem[{{Lin} {et~al.}(1996){Lin}, {Bodenheimer}, \&
  {Richardson}}]{1996Natur.380..606L}
{Lin}, D.~N.~C., {Bodenheimer}, P., \& {Richardson}, D.~C. 1996, \nat, 380,
  606, \dodoi{10.1038/380606a0}

\bibitem[{{Linsky} {et~al.}(2014){Linsky}, {Fontenla}, \&
  {France}}]{Linsky2014}
{Linsky}, J.~L., {Fontenla}, J., \& {France}, K. 2014, \apj, 780, 61,
  \dodoi{10.1088/0004-637X/780/1/61}

\bibitem[{{Lundkvist} {et~al.}(2016){Lundkvist}, {Kjeldsen}, {Albrecht},
  {Davies}, {Basu}, {Huber}, {Justesen}, {Karoff}, {Silva Aguirre}, {van
  Eylen}, {Vang}, {Arentoft}, {Barclay}, {Bedding}, {Campante}, {Chaplin},
  {Christensen-Dalsgaard}, {Elsworth}, {Gilliland}, {Handberg}, {Hekker},
  {Kawaler}, {Lund}, {Metcalfe}, {Miglio}, {Rowe}, {Stello}, {Tingley}, \&
  {White}}]{2016NatCo...711201L}
{Lundkvist}, M.~S., {Kjeldsen}, H., {Albrecht}, S., {et~al.} 2016, Nature
  Communications, 7, 11201, \dodoi{10.1038/ncomms11201}

\bibitem[{{Mamajek} \& {Bell}(2014)}]{Mamajek2014}
{Mamajek}, E.~E., \& {Bell}, C. P.~M. 2014, \mnras, 445, 2169,
  \dodoi{10.1093/mnras/stu1894}

\bibitem[{{Mann} {et~al.}(2016{\natexlab{a}}){Mann}, {Gaidos}, {Mace},
  {Johnson}, {Bowler}, {LaCourse}, {Jacobs}, {Vanderburg}, {Kraus}, {Kaplan},
  \& {Jaffe}}]{2016ApJ...818...46M}
{Mann}, A.~W., {Gaidos}, E., {Mace}, G.~N., {et~al.} 2016{\natexlab{a}}, \apj,
  818, 46, \dodoi{10.3847/0004-637X/818/1/46}

\bibitem[{{Mann} {et~al.}(2016{\natexlab{b}}){Mann}, {Newton}, {Rizzuto},
  {Irwin}, {Feiden}, {Gaidos}, {Mace}, {Kraus}, {James}, {Ansdell},
  {Charbonneau}, {Covey}, {Ireland}, {Jaffe}, {Johnson}, {Kidder}, \&
  {Vanderburg}}]{2016AJ....152...61M}
{Mann}, A.~W., {Newton}, E.~R., {Rizzuto}, A.~C., {et~al.} 2016{\natexlab{b}},
  \aj, 152, 61, \dodoi{10.3847/0004-6256/152/3/61}

\bibitem[{{Mann} {et~al.}(2018){Mann}, {Vanderburg}, {Rizzuto}, {Kraus},
  {Berlind}, {Bieryla}, {Calkins}, {Esquerdo}, {Latham}, {Mace}, {Morris},
  {Quinn}, {Sokal}, \& {Stefanik}}]{Mann2018}
{Mann}, A.~W., {Vanderburg}, A., {Rizzuto}, A.~C., {et~al.} 2018, \aj, 155, 4,
  \dodoi{10.3847/1538-3881/aa9791}

\bibitem[{{Mann} {et~al.}(2020){Mann}, {Johnson}, {Vanderburg}, {Kraus},
  {Rizzuto}, {Wood}, {Bush}, {Rockcliffe}, {Newton}, {Latham}, {Mamajek},
  {Zhou}, {Quinn}, {Thao}, {Benatti}, {Cosentino}, {Desidera}, {Harutyunyan},
  {Lovis}, {Mortier}, {Pepe}, {Poretti}, {Wilson}, {Kristiansen}, {Gagliano},
  {Jacobs}, {LaCourse}, {Omohundro}, {Schwengeler}, {Kane}, {Hill}, {Rabus},
  {Esquerdo}, {Berlind}, {Collins}, {Murawski}, {Aitken}, {Hazam Sallam},
  {Massey}, {Ricker}, {Vanderspek}, {Seager}, {Winn}, {Jenkins}, {Barclay},
  {Caldwell}, {Dragomir}, {Doty}, {Glidden}, {Tenenbaum}, {Torres}, {Twicken},
  \& {Villanueva}}]{Mann2020}
{Mann}, A.~W., {Johnson}, M.~C., {Vanderburg}, A., {et~al.} 2020, arXiv
  e-prints, arXiv:2005.00047.
\newblock \doarXiv{2005.00047}

\bibitem[{{McLean} {et~al.}(1998){McLean}, {Becklin}, {Bendiksen}, {Brims},
  {Canfield}, {Figer}, {Graham}, {Hare}, {Lacayanga}, {Larkin}, {Larson},
  {Levenson}, {Magnone}, {Teplitz}, \& {Wong}}]{McLean1998}
{McLean}, I.~S., {Becklin}, E.~E., {Bendiksen}, O., {et~al.} 1998, in Society
  of Photo-Optical Instrumentation Engineers (SPIE) Conference Series, Vol.
  3354, \procspie, ed. A.~M. {Fowler}, 566--578, \dodoi{10.1117/12.317283}

\bibitem[{{Monsignori Fossi} {et~al.}(1996){Monsignori Fossi}, {Landini}, {Del
  Zanna}, \& {Bowyer}}]{Monsignori1996}
{Monsignori Fossi}, B.~C., {Landini}, M., {Del Zanna}, G., \& {Bowyer}, S.
  1996, \apj, 466, 427, \dodoi{10.1086/177522}

\bibitem[{{Nagasawa} \& {Ida}(2011)}]{2011ApJ...742...72N}
{Nagasawa}, M., \& {Ida}, S. 2011, \apj, 742, 72,
  \dodoi{10.1088/0004-637X/742/2/72}

\bibitem[{{Newton} {et~al.}(2019){Newton}, {Mann}, {Tofflemire}, {Pearce},
  {Rizzuto}, {Vanderburg}, {Martinez}, {Wang}, {Ruffio}, {Kraus}, {Johnson},
  {Thao}, {Wood}, {Rampalli}, {Nielsen}, {Collins}, {Dragomir}, {Hellier},
  {Anderson}, {Barclay}, {Brown}, {Feiden}, {Hart}, {Isopi}, {Kielkopf},
  {Mallia}, {Nelson}, {Rodriguez}, {Stockdale}, {Waite}, {Wright}, {Lissauer},
  {Ricker}, {Vanderspek}, {Latham}, {Seager}, {Winn}, {Jenkins}, {Bouma},
  {Burke}, {Davies}, {Fausnaugh}, {Li}, {Morris}, {Mukai}, {Villase{\~n}or},
  {Villeneuva}, {De Rosa}, {Macintosh}, {Mengel}, {Okumura}, \&
  {Wittenmyer}}]{2019ApJ...880L..17N}
{Newton}, E.~R., {Mann}, A.~W., {Tofflemire}, B.~M., {et~al.} 2019, \apjl, 880,
  L17, \dodoi{10.3847/2041-8213/ab2988}

\bibitem[{{Noll} {et~al.}(2012){Noll}, {Kausch}, {Barden}, {Jones}, {Szyszka},
  {Kimeswenger}, \& {Vinther}}]{Noll2012}
{Noll}, S., {Kausch}, W., {Barden}, M., {et~al.} 2012, \aap, 543, A92,
  \dodoi{10.1051/0004-6361/201219040}

\bibitem[{{Oklop{\v c}i{\'c}} \& {Hirata}(2018)}]{Oklopovcic2018}
{Oklop{\v c}i{\'c}}, A., \& {Hirata}, C.~M. 2018, \apjl, 855, L11,
  \dodoi{10.3847/2041-8213/aaada9}

\bibitem[{{Owen}(2019)}]{Owen2019}
{Owen}, J.~E. 2019, Annual Review of Earth and Planetary Sciences, 47, 67,
  \dodoi{10.1146/annurev-earth-053018-060246}

\bibitem[{{Plavchan} {et~al.}(2009){Plavchan}, {Werner}, {Chen}, {Stapelfeldt},
  {Su}, {Stauffer}, \& {Song}}]{Plavchan2009}
{Plavchan}, P., {Werner}, M.~W., {Chen}, C.~H., {et~al.} 2009, \apj, 698, 1068,
  \dodoi{10.1088/0004-637X/698/2/1068}

\bibitem[{{Plavchan} {et~al.}(2020){Plavchan}, {Barclay}, {Gagn{\'e}}, {Gao},
  {Cale}, {Matzko}, {Dragomir}, {Quinn}, {Feliz}, {Stassun}, {Crossfield},
  {Berardo}, {Latham}, {Tieu}, {Anglada-Escud{\'e}}, {Ricker}, {Vanderspek},
  {Seager}, {Winn}, {Jenkins}, {Rinehart}, {Krishnamurthy}, {Dynes}, {Doty},
  {Adams}, {Afanasev}, {Beichman}, {Bottom}, {Bowler}, {Brinkworth}, {Brown},
  {Cancino}, {Ciardi}, {Clampin}, {Clark}, {Collins}, {Davison},
  {Foreman-Mackey}, {Furlan}, {Gaidos}, {Geneser}, {Giddens}, {Gilbert},
  {Hall}, {Hellier}, {Henry}, {Horner}, {Howard}, {Huang}, {Huber}, {Kane},
  {Kenworthy}, {Kielkopf}, {Kipping}, {Klenke}, {Kruse}, {Latouf}, {Lowrance},
  {Mennesson}, {Mengel}, {Mills}, {Morton}, {Narita}, {Newton}, {Nishimoto},
  {Okumura}, {Palle}, {Pepper}, {Quintana}, {Roberge}, {Roccatagliata},
  {Schlieder}, {Tanner}, {Teske}, {Tinney}, {Vanderburg}, {von Braun}, {Walp},
  {Wang}, {Wang}, {Weigand }, {White}, {Wittenmyer}, {Wright}, {Youngblood},
  {Zhang}, \& {Zilberman}}]{2020Natur.582..497P}
{Plavchan}, P., {Barclay}, T., {Gagn{\'e}}, J., {et~al.} 2020, \nat, 582, 497,
  \dodoi{10.1038/s41586-020-2400-z}

\bibitem[{{Price-Whelan} {et~al.}(2018){Price-Whelan}, {Sip{\H{o}}cz},
  {G{\"u}nther}, {Lim}, {Crawford}, {Conseil}, {Shupe}, {Craig}, {Dencheva},
  {Ginsburg}, {VanderPlas}, {Bradley}, {P{\'e}rez-Su{\'a}rez}, {de Val-Borro},
  {Paper Contributors}, {Aldcroft}, {Cruz}, {Robitaille}, {Tollerud},
  {Coordination Committee}, {Ardelean}, {Babej}, {Bach}, {Bachetti}, {Bakanov},
  {Bamford}, {Barentsen}, {Barmby}, {Baumbach}, {Berry}, {Biscani}, {Boquien},
  {Bostroem}, {Bouma}, {Brammer}, {Bray}, {Breytenbach}, {Buddelmeijer},
  {Burke}, {Calderone}, {Cano Rodr{\'\i}guez}, {Cara}, {Cardoso}, {Cheedella},
  {Copin}, {Corrales}, {Crichton}, {D{\textquoteright}Avella}, {Deil},
  {Depagne}, {Dietrich}, {Donath}, {Droettboom}, {Earl}, {Erben}, {Fabbro},
  {Ferreira}, {Finethy}, {Fox}, {Garrison}, {Gibbons}, {Goldstein}, {Gommers},
  {Greco}, {Greenfield}, {Groener}, {Grollier}, {Hagen}, {Hirst}, {Homeier},
  {Horton}, {Hosseinzadeh}, {Hu}, {Hunkeler}, {Ivezi{\'c}}, {Jain}, {Jenness},
  {Kanarek}, {Kendrew}, {Kern}, {Kerzendorf}, {Khvalko}, {King}, {Kirkby},
  {Kulkarni}, {Kumar}, {Lee}, {Lenz}, {Littlefair}, {Ma}, {Macleod},
  {Mastropietro}, {McCully}, {Montagnac}, {Morris}, {Mueller}, {Mumford},
  {Muna}, {Murphy}, {Nelson}, {Nguyen}, {Ninan}, {N{\"o}the}, {Ogaz}, {Oh},
  {Parejko}, {Parley}, {Pascual}, {Patil}, {Patil}, {Plunkett}, {Prochaska},
  {Rastogi}, {Reddy Janga}, {Sabater}, {Sakurikar}, {Seifert}, {Sherbert},
  {Sherwood-Taylor}, {Shih}, {Sick}, {Silbiger}, {Singanamalla}, {Singer},
  {Sladen}, {Sooley}, {Sornarajah}, {Streicher}, {Teuben}, {Thomas},
  {Tremblay}, {Turner}, {Terr{\'o}n}, {van Kerkwijk}, {de la Vega}, {Watkins},
  {Weaver}, {Whitmore}, {Woillez}, {Zabalza}, \& {Contributors}}]{astropy:2018}
{Price-Whelan}, A.~M., {Sip{\H{o}}cz}, B.~M., {G{\"u}nther}, H.~M., {et~al.}
  2018, \aj, 156, 123, \dodoi{10.3847/1538-3881/aabc4f}

\bibitem[{{Quinn} {et~al.}(2012){Quinn}, {White}, {Latham}, {Buchhave},
  {Cantrell}, {Dahm}, {F{\H{u}}r{\'e}sz}, {Szentgyorgyi}, {Geary}, {Torres},
  {Bieryla}, {Berlind}, {Calkins}, {Esquerdo}, \&
  {Stefanik}}]{2012ApJ...756L..33Q}
{Quinn}, S.~N., {White}, R.~J., {Latham}, D.~W., {et~al.} 2012, \apjl, 756,
  L33, \dodoi{10.1088/2041-8205/756/2/L33}

\bibitem[{{Robinson} {et~al.}(2001){Robinson}, {Linsky}, {Woodgate}, \&
  {Timothy}}]{Robinson2001}
{Robinson}, R.~D., {Linsky}, J.~L., {Woodgate}, B.~E., \& {Timothy}, J.~G.
  2001, \apj, 554, 368, \dodoi{10.1086/321379}

\bibitem[{{Sato} {et~al.}(2007){Sato}, {Izumiura}, {Toyota}, {Kambe}, {Takeda},
  {Masuda}, {Omiya}, {Murata}, {Itoh}, {Ando}, {Yoshida}, {Ikoma}, {Kokubo}, \&
  {Ida}}]{2007ApJ...661..527S}
{Sato}, B., {Izumiura}, H., {Toyota}, E., {et~al.} 2007, \apj, 661, 527,
  \dodoi{10.1086/513503}

\bibitem[{{Schneider} \& {Shkolnik}(2018)}]{Schneider2018}
{Schneider}, A.~C., \& {Shkolnik}, E.~L. 2018, \aj, 155, 122,
  \dodoi{10.3847/1538-3881/aaaa24}

\bibitem[{{Shematovich} {et~al.}(2014){Shematovich}, {Ionov}, \&
  {Lammer}}]{Shematovich2014}
{Shematovich}, V.~I., {Ionov}, D.~E., \& {Lammer}, H. 2014, \aap, 571, A94,
  \dodoi{10.1051/0004-6361/201423573}

\bibitem[{{Smith}(2016)}]{Smith2016}
{Smith}, G.~H. 2016, PASA, 33, e057, \dodoi{10.1017/pasa.2016.49}

\bibitem[{{Spalding} \& {Batygin}(2016)}]{Spalding2016}
{Spalding}, C., \& {Batygin}, K. 2016, \apj, 830, 5,
  \dodoi{10.3847/0004-637X/830/1/5}

\bibitem[{{Tamura} {et~al.}(2012){Tamura}, {Suto}, {Nishikawa}, {Kotani},
  {Sato}, {Aoki}, {Usuda}, {Kurokawa}, {Kashiwagi}, {Nishiyama}, {Ikeda},
  {Hall}, {Hodapp}, {Hashimoto}, {Morino}, {Inoue}, {Mizuno}, {Washizaki},
  {Tanaka}, {Suzuki}, {Kwon}, {Suenaga}, {Oh}, {Narita}, {Kokubo}, {Hayano},
  {Izumiura}, {Kambe}, {Kudo}, {Kusakabe}, {Ikoma}, {Hori}, {Omiya}, {Genda},
  {Fukui}, {Fujii}, {Guyon}, {Harakawa}, {Hayashi}, {Hidai}, {Hirano},
  {Kuzuhara}, {Machida}, {Matsuo}, {Nagata}, {Ohnuki}, {Ogihara}, {Oshino},
  {Suzuki}, {Takami}, {Takato}, {Takahashi}, {Tachinami}, \&
  {Terada}}]{2012SPIE.8446E..1TT}
{Tamura}, M., {Suto}, H., {Nishikawa}, J., {et~al.} 2012, in \procspie, Vol.
  8446, Ground-based and Airborne Instrumentation for Astronomy IV, 84461T,
  \dodoi{10.1117/12.925885}

\bibitem[{{Tody}(1993)}]{1993ASPC...52..173T}
{Tody}, D. 1993, in Astronomical Society of the Pacific Conference Series,
  Vol.~52, Astronomical Data Analysis Software and Systems II, ed. R.~J.
  {Hanisch}, R.~J.~V. {Brissenden}, \& J.~{Barnes}, 173

\bibitem[{{Vazan} {et~al.}(2018){Vazan}, {Ormel}, \& {Dominik}}]{Vazan2018}
{Vazan}, A., {Ormel}, C.~W., \& {Dominik}, C. 2018, \aap, 610, L1,
  \dodoi{10.1051/0004-6361/201732200}

\bibitem[{{Winn} {et~al.}(2010){Winn}, {Fabrycky}, {Albrecht}, \&
  {Johnson}}]{2010ApJ...718L.145W}
{Winn}, J.~N., {Fabrycky}, D., {Albrecht}, S., \& {Johnson}, J.~A. 2010, \apjl,
  718, L145, \dodoi{10.1088/2041-8205/718/2/L145}

\bibitem[{{Winn} {et~al.}(2005){Winn}, {Noyes}, {Holman}, {Charbonneau},
  {Ohta}, {Taruya}, {Suto}, {Narita}, {Turner}, {Johnson}, {Marcy}, {Butler},
  \& {Vogt}}]{2005ApJ...631.1215W}
{Winn}, J.~N., {Noyes}, R.~W., {Holman}, M.~J., {et~al.} 2005, \apj, 631, 1215,
  \dodoi{10.1086/432571}

\end{thebibliography}
\bibliographystyle{aasjournal}

%% This command is needed to show the entire author+affiliation list when
%% the collaboration and author truncation commands are used.  It has to
%% go at the end of the manuscript.
%\allauthors

%% Include this line if you are using the \added, \replaced, \deleted
%% commands to see a summary list of all changes at the end of the article.
%\listofchanges

\end{document}